\documentclass[letter]{aa}     

\usepackage{graphicx}
\usepackage{txfonts}
\usepackage{lipsum}
\usepackage{subcaption}         
\usepackage{lscape}             
\usepackage{placeins}           
                                
\usepackage{color}
	\definecolor{unipd}{HTML}{b5121b}
	\definecolor{cadmiumgreen}{RGB}{2,134,121}
	\definecolor{blue}{RGB}{0,32,169}
\usepackage{hyperref}
   \hypersetup{colorlinks=true, linkbordercolor=blue, breaklinks=true, citecolor=blue, filecolor=black, linkcolor=blue, urlcolor=cadmiumgreen}

\begin{document}

   \title{Mergers via failed common envelope as a route towards intermediate-mass stripped stars}



   \author{A. Picco
          \inst{1,3}
          \and
          P. Marchant\inst{2,1}
          \and
          D. Pauli\inst{1}
          \and
          H. Sana\inst{1,3}
          }

   \institute{Institute of Astronomy, KU Leuven, Celestijnlaan 200D, 3001 Leuven,
        Belgium,\\
              \email{annachiara.picco@kuleuven.be}
         \and
         Sterrenkundig Observatorium, Universiteit Gent, Krijgslaan 281 S9, B-9000 Gent, Belgium,
         \and
         Leuven Gravity Institute, KU Leuven, Celestijnenlaan 200D, box 2415, 3001 Leuven, Belgium
             }

   \date{\today}

 
  \abstract
   {Stripped stars are a common product of binary stellar systems and span a wide mass range from Wolf-Rayet stars to hot subdwarfs and helium white dwarfs. The recent discovery of intermediate-mass stripped stars, with masses between those of WRs and subdwarfs, provides a continuous evolutionary sequence and valuable test beds for binary interaction models. Population synthesis studies explain the formation of stripped products through stable mass transfer or common envelope evolution. Recently, however, a merger scenario (or, "failed common envelope") between a stripped star and a post-main sequence companion was suggested to explain the formation of the magnetic intermediate-mass stripped star HD 45166. In this Letter, we investigate this channel using detailed interacting binary simulations at Galactic metallicity and model their stripped merger products. We find that these merger products can contribute to the population of stripped stars across the Hertzsprung-Russell diagram and produce a blue-straggler effect, with merger masses spanning from $0.6\,M_{\odot}$ to $14\,M_{\odot}$. These objects would be consistent with long-lived core He-burning stars retaining thin hydrogen envelopes after a partial CE ejection, and they would appear as single or members of wide binaries that were originally hierarchical triples.}

   \keywords{binaries: close -- binaries: evolution -- stars: evolution -- stars: mass-loss -- stars: Wolf-Rayet -- stars: subdwarfs}

   \maketitle
   \nolinenumbers

\section{Introduction}

\vspace{-0.1cm}
Stripped stars are stars that have lost their hydrogen-rich envelopes to a companion, and they are a common outcome of the evolution of binary systems \citep{Sana_2012, Sana2025}.
The stripping phenomenon in binaries occurs at any stellar mass, but the spectral appearance of the resulting stripped star strongly depends on its mass and mass-loss properties. At the high-mass end, stripping can produce Wolf–Rayet (WR) stars of mass $M \gtrsim 8\:M_{\odot}$ \citep{crowther_physical_2007, Shenar2020}; at the low-mass end, it produces OB-type subdwarfs (sdOBs; \citealt{heberHotSubdwarfStars2025}) and helium (He) white dwarfs (\citealt{Brown2020}) of mass $M \lesssim 2\:M_{\odot}$. The recent discovery of intermediate-mass ($2 \lesssim M/M_{\odot} \lesssim 8$) stripped stars  \citealt{droutDiscoveryMissingIntermediatemass2023}) has completed the observational bridge between the low- and high-mass stripped products.

The formation of stripped stars has been studied in the low-mass regime, where population synthesis models (\citealt{Han2002}) can reproduce the observed properties of sdOBs through two evolutionary channels that remove the envelope: stable mass transfer (MT) via Roche-lobe overflow (RLOF; \citealt{kippenhahn_entwicklung_1967}) and common envelope (CE) evolution \citep{Paczynski1971}. At higher masses, the same MT channels may produce both intermediate-mass stripped stars (\citealt{Shao2021}, \citealt{Yungelson2024}, \citealt{HovisAfflerbach2025}) and WR stars \citep{vanbeveren1998}, which can also form in isolation via stellar winds in the Conti scenario \citep{Conti1975}. 
For intermediate-mass stripped stars, the sample is still limited to a few spectroscopically confirmed objects (see \citealt{gotberg_stellar_2023} for the Magellanic Clouds; HD 45166 \citealt{Shenar2023} in the Galaxy), and recent UV surveys identiﬁed 820 more candidates (\citealt{ludwigStrippedStarUltravioletMagellanic2025}).

The idea that two stripped stars may merge to form a more massive stripped product has been invoked for sdBs. In particular, the merger of two He white dwarfs driven by gravitational-wave emission represents a third formation channel (\citealt{Webbink1984}), in addition to stable RLOF and CE evolution. Observationally, this scenario may explain a third of the known single sdBs (\citealt{heberHotSubdwarfStars2025}) and the growing sample of massive ($\sim 0.6\:M_{\odot}$) He sdOs (\citealt{Latour2025}).
Similarly, mergers have been suggested as an origin for WR stars, in the case where hydrogen-rich massive merger products undergo self-stripping via wind-driven mass loss. These may account for up to $\sim60\%$ of the WR population in the Milky Way \citep{Li2024}, consistent with the sample of \citet{Deshmukh2024}. Such a channel may also aid in explaining single WR stars in low metallicity environments (\citealt{Schootemeijer2024}, \citealt{Gilkis2025}).

\begin{figure}[t]
   \centering
   \includegraphics[width=0.8\hsize]{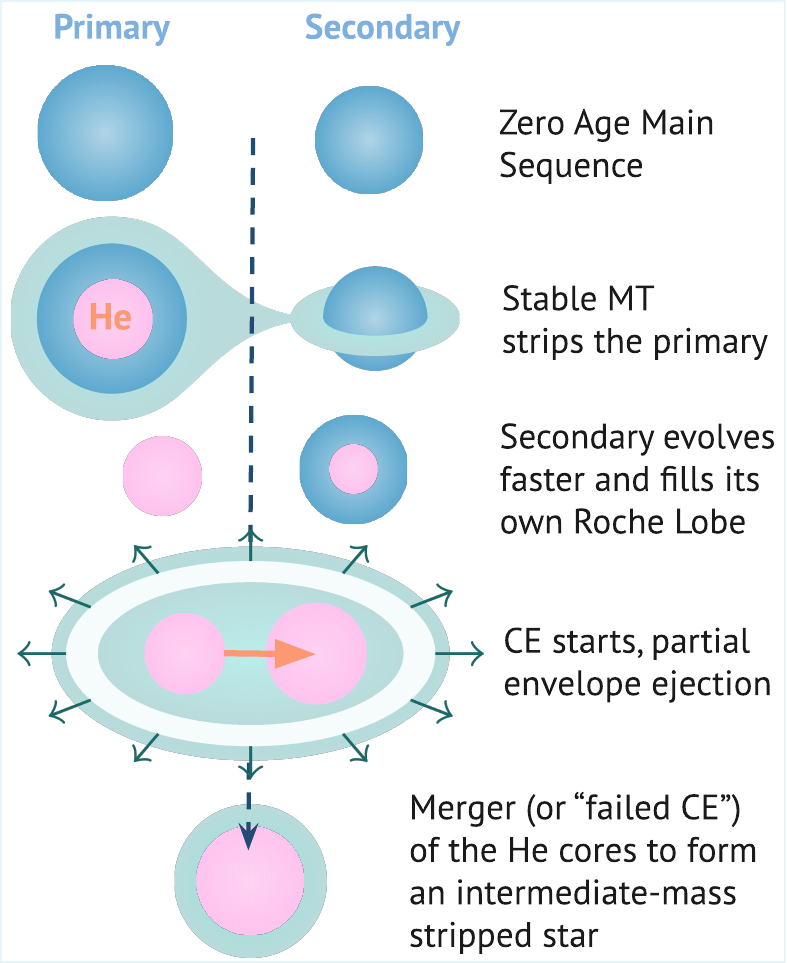}
      \caption{Sketch of the merger via failed CE channel towards intermediate-mass He-stars.}
         \label{fig:diagram}
   \end{figure}

An especially interesting case is HD~45166, a so-called quasi-WR star due to its spectral appearance (\citealt{Steiner2005}). Interpreted originally as the first example of an intermediate-mass stripped star in the Galaxy, it was then also proposed to have formed through the merger (or, "failed CE") of a stripped He star (the core of the initially more massive component) with a hydrogen-rich post-MS star (the accretor during the first stable RLOF) (\citealt{Shenar2023}). The energy released through this merger would eject almost the entire hydrogen envelope, leaving an isolated stripped star. So far, no population synthesis study has considered such mergers as a possible way of forming stripped stars. This Letter builds upon the evolutionary interpretation of HD 45166 and explores the failed-CE mergers scenario (sketched in Fig. \ref{fig:diagram}) in the context of the formation of galactic intermediate-mass stripped stars. 

\vspace{-0.4cm}
\section{Methodology}
We use version 24.03.1 of Modules for Experiments in Stellar Astrophysics
\citep[MESA][]{Paxton2011, Paxton2015, Jermyn2023}. Details about our setup are provided in Appendix \ref{sec:appMESA} and the input files are uploaded to Zenodo\footnote{\href{https://doi.org/10.5281/zenodo.17424084}{doi.org/10.5281/zenodo.17424084}}.

\begin{figure}[t]
   \centering
   \includegraphics[width=\hsize]{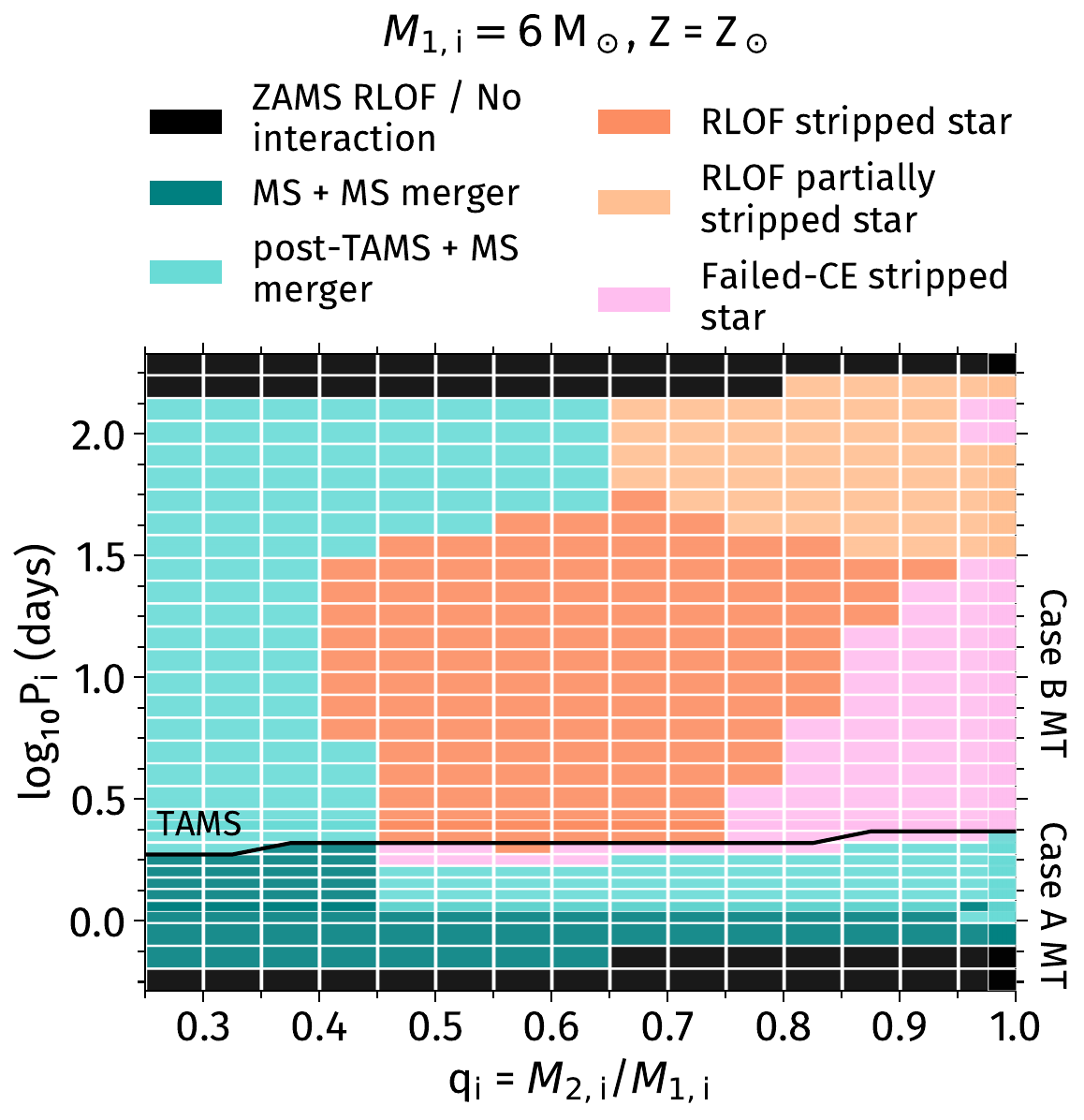}
      \caption{Example of one of our MESA grids for a primary star with mass $M_{1,\:\mathrm{i}}=6\:M_{\odot}$ at solar metallicity, spanning initial mass ratios $q_{\mathrm{i}}\equiv M_{2,\:\mathrm{i}}/M_{1,\:\mathrm{i}}$ and orbital periods $P_{\mathrm{i}}$. The resolution is finer around the TAMS (black line) and around $q_{\mathrm{i}}=1$. Different outcomes are color-coded as in the legend: post-TAMS (MS) + MS merger refer to mergers between a post-TAMS (MS) star with a MS companion. RLOF stripped stars are formed after stable MT from the primary and detachment, and we refer to RLOF partial stripping when the He core is $<\:90\%$ of the total mass. }
         \label{fig:grid}
   \end{figure}

We first produced single-star tracks at solar metallicity, $Z_{\odot}$, at initial masses $2 < M_{1,\:\mathrm{i}}/ M_{\odot} < 20$ in steps of $\Delta M_{1,\:\mathrm{i}} = 1\:M_{\odot}$, until core He depletion. We refer to the primary (secondary) star with mass $M_{1,\:\mathrm{i}}$ ($M_{2,\:\mathrm{i}}$) as the initially more (less) massive component. From the single-star tracks we extract the minimum and maximum orbital periods, $\log P_{\mathrm{min}}$ and $\log P_{\mathrm{max}}$ (days), corresponding to binaries in which the primary undergoes RLOF at the Zero-Age Main Sequence (ZAMS) and in which no interaction occurs up to He depletion, respectively, for a given initial mass ratio $q_{\mathrm{i}}\equiv M_{2,\:\mathrm{i}}/M_{1,\:\mathrm{i}}$. For each $M_{1,\:\mathrm{i}}$ we computed a grid of binary models spanning $0.275 \leq q_{\mathrm{i}} \leq 1$ with spacing $\Delta q_{\mathrm{i}} = 0.05$, and a variable resolution in orbital period across $\log P_{\mathrm{min}} \leq \log P\:(\mathrm{days}) \leq \log P_{\mathrm{max}}$ such that each grid spans 28 period values. We additionally refine the sampling around the Terminal-Age Main Sequence (TAMS) and around $q_{\mathrm{i}} = 1$ to resolve the parameter space relevant for the failed CE mergers (see below). An example grid for $M_{1,\:\mathrm{i}} = 6\:M_{\odot}$ is shown in Fig.~\ref{fig:grid}.

We flag a system as stripped star produced by a failed CE when both of the following conditions are met: 1) the binary merges in a CE episode; 2) at the onset of such CE episode, both stars have evolved past their TAMS, and one star has already been stripped by a (stable) MT episode before (as in Fig. \ref{fig:diagram}). We model CE evolution using a double-core energy formalism where both stellar components contribute to the envelope ejection and orbital shrinkage. The CE mass-loss rate is regulated by the degree of Roche-lobe overfilling of both the donor and accretor, allowing mass loss to persist until both stars detach or merge. Full details about our definition of CE onset and the CE treatment are provided in Appendix \ref{sec:appMethods}.

%

\vspace{-0.4cm}
\section{Results}
\begin{figure}[t]
   \centering
   \includegraphics[width=\hsize]{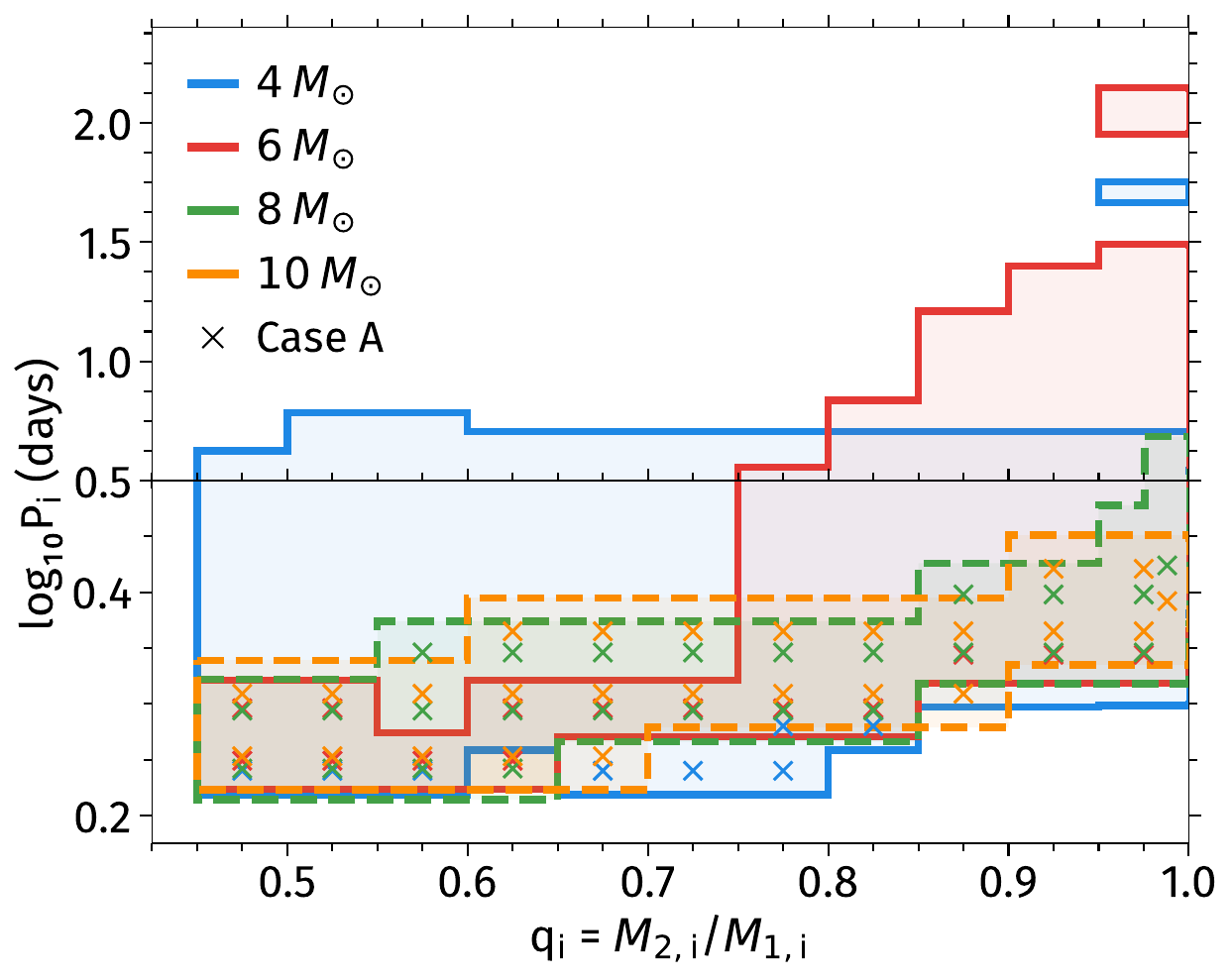}
      \caption{Parameter space in initial period $P_\mathrm{i}$ and mass ratio $q_\mathrm{i}$ for failed-CE stripped stars for a selection of values of initial primary masses $M_{1,\:\mathrm{i}}$ (color-coded). The part of the parameter space that undergoes case A MT in the first interaction is also indicated with x scatter points. Full grids similar to Fig. \ref{fig:grid} but for the considered $M_{1,\:\mathrm{i}}$ are shown in Appendix \ref{sec:appGrids}.}
         \label{fig:shapes}
   \end{figure}

The parameter space in initial orbital period $P_{\mathrm{i}}$ and mass ratio $q_{\mathrm{i}}$ leading to failed-CE stripped stars depends on the initial primary mass $M_{1,\mathrm{i}}$, as shown in Fig. \ref{fig:shapes} (see also Appendix \ref{sec:appGrids}). The channel is present for all primary masses, but the allowed parameter space is increasingly suppressed for $M_{1,\mathrm{i}} > 10\,M_\odot$ and disappears entirely (at our resolution) for $M_{1,\mathrm{i}} > 18\,M_\odot$ (see Appendix \ref{sec:appGrids}); additionally, we see a suppression for $M_{1,\:\mathrm{i}}<3\:M_{\odot}$.

At any $M_{1,\mathrm{i}}$, systems undergoing a first case A stable MT episode from the primary may result in failed-CE mergers even for extreme initial mass ratios. These are mainly binaries in which the primary undergoes stable case AB MT and, while it burns He in its core as a detached stripped star, the secondary evolves past-TAMS and initiates the failed CE (scenario of Fig. \ref{fig:diagram}). In a few systems, the secondary completes core-hydrogen burning and triggers the merger during the case AB MT. Binaries that form failed-CE stripped stars after a first case B stable MT episode, instead, always follow the evolutionary scenario depicted in Fig. \ref{fig:diagram}, and form stripped stars through failed CE at initial mass ratios closer and closer to unity, as $M_{1,\mathrm{i}}$ increases. 

In general, we expect a region below the TAMS to always favor the failed-CE stripped star outcome, due to the tight initial orbits and the possibility to strip the primary down to its not fully formed He core, leading to an extreme post-MT mass ratio. Additionally, we expect mass ratios $q_{\mathrm{i}} \simeq 1$ to favor the merger outcome, as the stars evolve at nearly the same pace and a small amount of MT is needed to invert the mass ratio. Our resolution, however, does not resolve the narrow region around $q_{\mathrm{i}} \simeq 1$ for $M_{1,\mathrm{i}} \geq 10\,M_\odot$. Similarly, despite our increased resolution around TAMS, we do not see the failed-CE stripped star region for $M_{1,\mathrm{i}} \geq 18\,M_\odot$. Even if one would resolve the region with a better sampling of the initial conditions, this is indicative of a fine-tuning of the channel at higher masses.

For case B systems, the failed-CE stripped star outcome is inhibited at longer initial orbital periods, with a dependence on how close the mass ratio is to unity (see, e.g., the triangular shape above TAMS in Fig. \ref{fig:grid}, and Appendix \ref{sec:appContours}). This dependence is present at all primary masses, but with decreasing importance at lower $M_{1,\mathrm{i}}$ (see, e.g., the semi-rectangular shape of $M_{1,\mathrm{i}}=4\:M_{\odot}$ in Fig. \ref{fig:shapes}), due to the combination of two effects: lower mass stars 1) have less massive He cores relative to their total mass, so they can transfer more mass during MT and produce more extreme post-MT mass ratios; 2) have a steeper mass-luminosity relation, such that the secondary’s evolutionary acceleration due to accretion is favored. We can see these effects in action for primary masses $M_{1,\mathrm{i}} < 6\,M_\odot$, where case B systems can form stripped stars via failed-CE for initial mass ratios farther from unity than for more massive primaries (see also Appendix \ref{sec:appContours}). 


%

\section{Discussion}

\begin{figure}[t]
   \centering
   \includegraphics[width=\hsize]{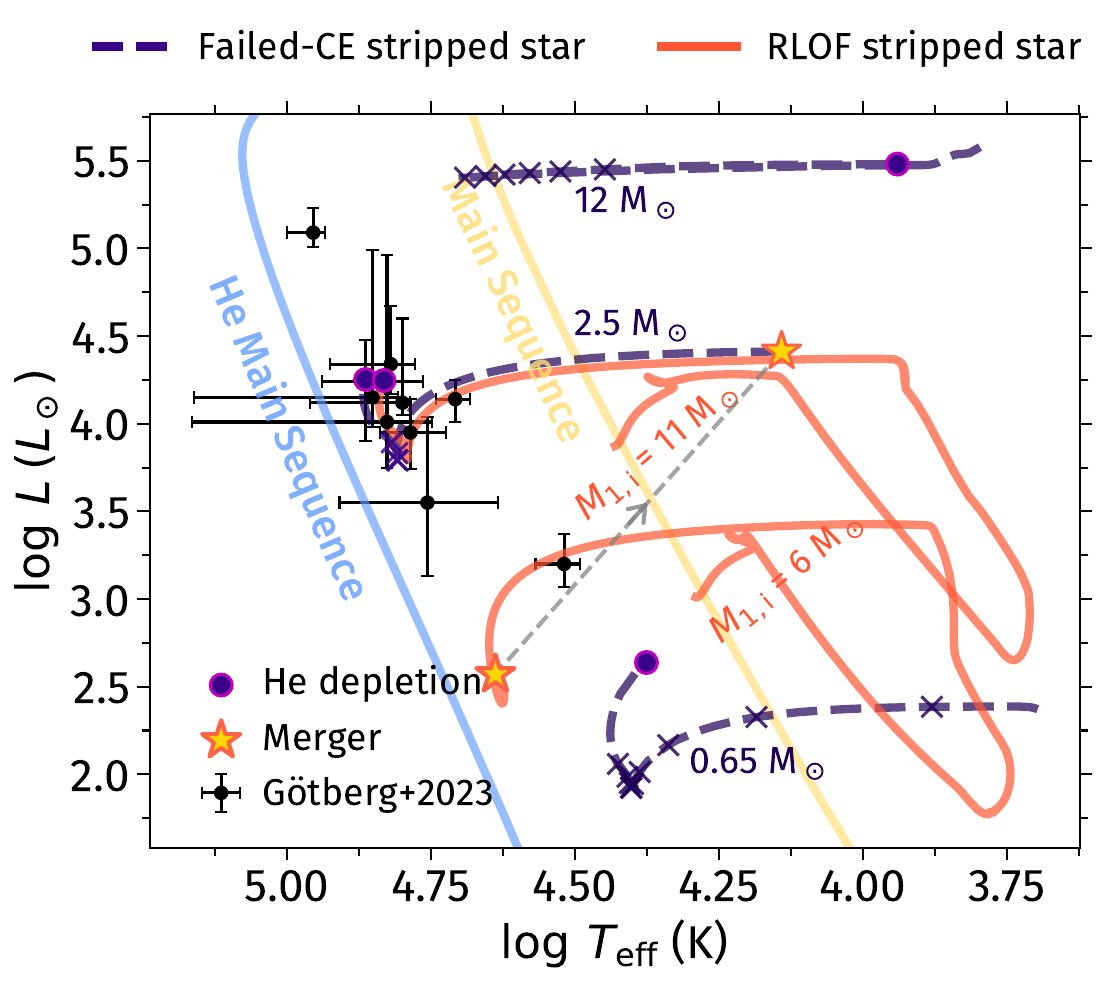}
      \caption{Location of the stripped star sample of \citealt{gotberg_stellar_2023} in the HR diagram compared to a selection of our failed-CE stripped star models. For the 2.5 $M_{\odot}$ we show the evolution of its 6 $M_{\odot}$ primary prior to the failed CE (pre- and post-merger stages connected through an arrow). We also show the evolution of a 11 $M_{\odot}$ donor star which produces a 2.5 $M_{\odot}$ stripped star through RLOF. Cross markers for the merger models are equally spaced by 5, 0.5 and 0.05 Myr for the 0.65, 2.5 and 12 $M_{\odot}$ stripped star models, respectively. In Appendix \ref{sec:appHR} we show the complete evolution of the represented tracks. We also show a theoretical Main Sequence and He Main Sequence in blue and yellow.}
         \label{fig:HR} 
   \end{figure}

The failed-CE merger channel is robust across all the considered masses. At our resolution, the minimum and maximum merger-product masses are $0.6\:M_{\odot}$ and $14\:M_{\odot}$, respectively. We expect that higher mass stripped products may be produced for $M_{1,\mathrm{i}}\geq 18\:M_{\odot}$, at fine-tuned mass ratios (around unity or close to TAMS) that we do not resolve. This shows that the channel may contribute to the population of stripped products across the HR diagram, from high-mass He subdwarfs \citep{Latour2025} to stripped stars showing the WR phenomenon at solar metallicity \citep{SanderVink2020}. We expect the population of failed-CE stripped stars to produce a blue-straggler effect, due to the creation of higher-mass (and therefore higher-luminosity) interaction products than those expected to accumulate at a cluster's turn-off from standard RLOF stripping. This is illustrated in Fig. \ref{fig:HR} for an example case of a 2.5 $M_{\odot}$ failed-CE stripped star, which populates the HR diagram at a much higher luminosity than its 6 $M_{\odot}$ star progenitor; such a massive stripped product may be produced by simple RLOF stripping only from a much more massive progenitor, with initial mass 11 $M_{\odot}$ (see also Appendix \ref{sec:appHR}). Observationally, the failed-CE stripped stars are expected to appear in isolation or as members of wide binaries that were originally hierarchical triples, and may show signatures of magnetism, as in the case of HD~45166 \citep{Shenar2023}. As shown by \citet{Shenar2023}, the effective temperature reached by such products (and, therefore, their proximity to the sample of \citet{gotberg_stellar_2023}, see Fig. \ref{fig:HR}) depends on the assumed CE ejection efficiency, which we do not explore in this study.

%

\begin{figure}[t]
   \centering
   \includegraphics[width=\hsize]{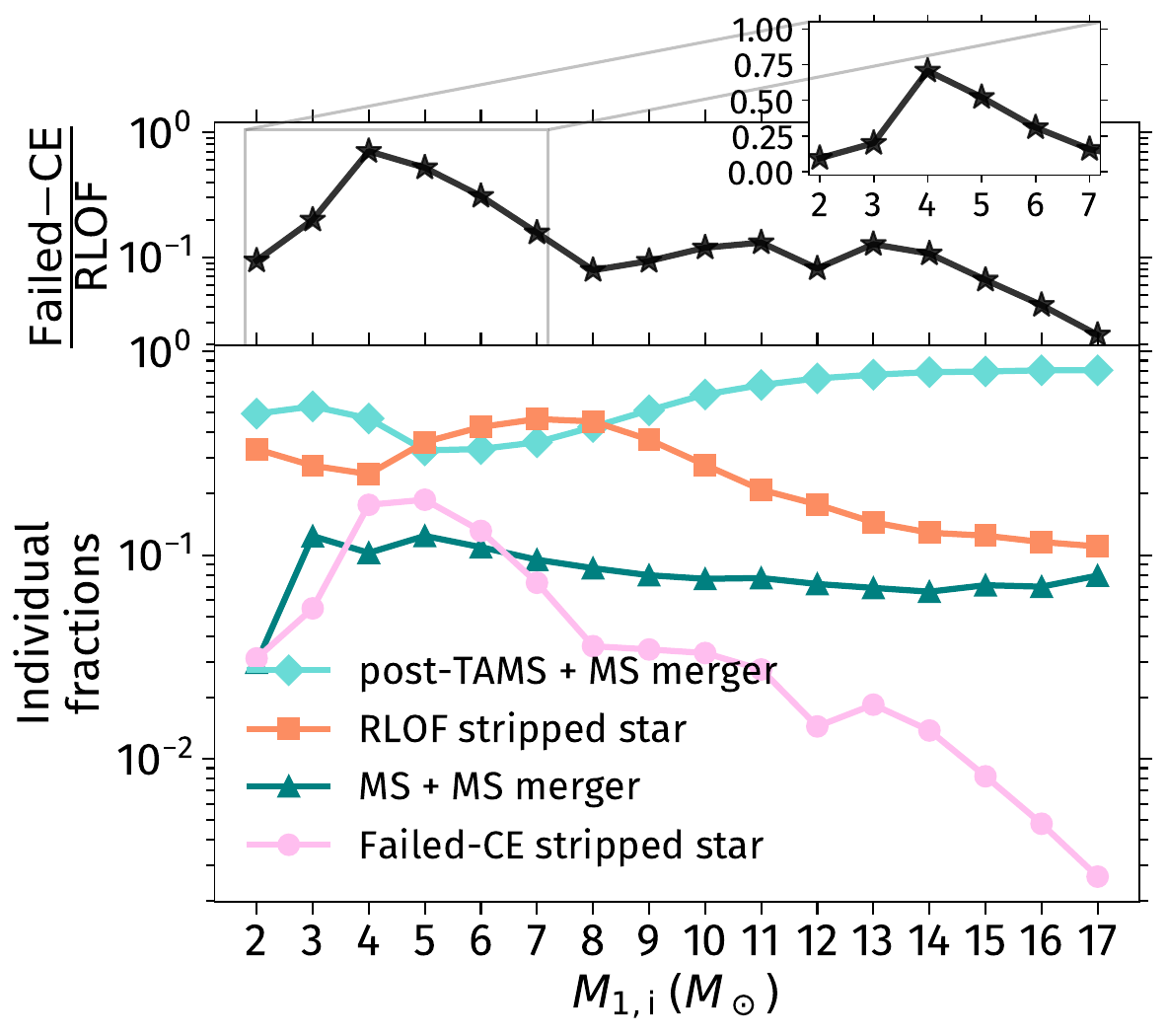}
      \caption{Bottom panel: fractions of systems with different outcomes (color coded) with respect to the total area of interacting binaries for each of our grids at different primary mass $M_{1,\:\mathrm{i}}$, as a function of $M_{1,\:\mathrm{i}}$. Top panel: relative fraction of systems with outcome failed-CE versus those with RLOF stripped star.}
         \label{fig:rates}
   \end{figure}

A full population synthesis study is needed to assess the effective contribution of this channel to the overall population of stripped stars. Nevertheless, we can already provide an estimate of the likelihood of a binary with a given initial primary mass to produce a failed-CE stripped star relative to that of producing other binary interaction outcomes (Fig.~\ref{fig:rates}), assuming uniform distribution in $\log P_{\mathrm{i}}$ and $q_{\mathrm{i}}$. We find that the failed-CE merger products can constitute between 6 and 20\% of the outcomes of interacting binaries for primary masses $3<M_{1,\:\mathrm{i}}\:(M_{\odot})<7$. Restricting to systems that form stripped stars, we find that the failed-CE outcome is less likely than regular RLOF stripping at all primary masses, but their formation rate can still be comparable; at 4 $M_{\odot}$, in particular, the ratio of systems producing stripped stars through failed CE versus those with stable RLOF peaks at 75\% (see inset of Fig. \ref{fig:rates}). A population synthesis analysis should, however, account for evolutionary timescales and the birth distributions of binary parameters. On the one hand, failed-CE stripped stars undergo a RLOF stripped-star phase before merging; on the other hand, this channel is favoured at shorter initial orbital periods, at which high-mass binaries are more likely to exist (\citealt{Sana2025}). In addition, our models are terminated at core He depletion, but post-He-depletion expansion \citep{Laplace2020} may trigger a late MT episode, or the rejuvenated secondary may initiate a late CE phase and a possible merger. These systems would host inert carbon-oxygen cores and are expected to be short-lived, and therefore unlikely to make a significant contribution to the population. Additionally, there is a number of unknowns in our results that we do not explore, the most important ones being the effect of metallicity, the assumption on the overshooting parameter, and the MT efficiency. Initial experiments, which we discuss in Appendix~\ref{sec:appVariations}, indicate that the failed-CE channel may be suppressed for lower metallicities and higher overshooting, as well as at lower MT efficiency. We defer the treatment of these effects to future work.

The hydrogen envelope left behind by a successful CE ejection determines the excursion in the HR diagram and the observational properties of the merger products \citep{Shenar2023}; mass loss during a failed CE can also shape the resulting supernova type \citep{Zapartas2019}. Our models compute self-consistently the remaining post-merger hydrogen layer from the binding energy of the expelled envelopes of both stars (see also Appendix~\ref{sec:appMethods}). This represents an improvement over standard binary population synthesis frameworks, which typically assume either complete envelope removal or rely on crude approximations based only on the donor's envelope. Although our model is self-consistent, it still relies on a free parameter describing the CE efficiency. Nevertheless, we expect our results to provide a lower limit on the amount of envelope removed, as we consider fully efficient CE and ignore any additional energy resulting from the coalescence of the cores (see Appendix \ref{sec:appMethods}). Tailored multi-dimensional hydrodynamical simulations are critical to assess the goodness of our assumptions.

%

\vspace{-0.4cm}
\section{Conclusions}
We find that failed-CE stripped stars may populate regions of the HR diagram where intermediate-mass stripped stars are expected. These products would appear as single or members of wide binaries that were originally hierarchical triples, and produce a blue-straggler effect. Their post-merger evolution strongly depends on the remaining hydrogen envelope mass, for which we compute a self-consistent lower bound within one-dimensional CE simulations, but this scenario ultimately calls out for dedicated multi-dimensional hydrodynamical studies.

\begin{acknowledgements}
The authors thank the referee Yong Shao for his useful comments; Mathias Mertens, Emma Zoe Casier and Sébastien Kinif for the engagement in the early stage of the project; Max Briel and Anastasios Fragkos for the interesting discussions. AP acknowledges support from the Research Foundation - Flanders (FWO), grant agreement No. 11M8325N (PhD Fellowship), and the European Union’s Horizon (ERC) Europe programme under grant agreement No 101131928, project ACME. PM acknowledges support from the ERC (European Union’s Horizon 2020 research and innovation programme) under grant No. 101165213/Star-Grasp, and FWO postdoctoral fellowship number 12ZY523N. DP acknowledges financial support from the FWO postdoctoral fellowship No. 1256225N. The resources used in this work were provided by the Flemish Supercomputer Center, funded by the FWO and the Flemish Government.
\end{acknowledgements}

%

\bibliographystyle{aa}
\vspace{-0.4cm}
\bibliography{My_Library.bib}

\begin{appendix}
\section{Details of \text{MESA} simulations}\label{sec:appMESA}
Our simulations are computed using version 24.03.1 of \text{MESA}. Our setup is builds upon that of \cite{Marchant2021}, to which we refer for more details. MT rates are computed taking into account MT through the first Lagrangian point $L_1$ from an extended atmosphere and optically thick regions. Notably, mass and angular momentum loss from the outer Langrangian point (L2) of the donor is not calculated in our setup (see also Appendix \ref{sec:appMethods}). We use such a modified 'Kolb' scheme (\citealt{KolbRitter1990}) because we model a mass loss rate from both stars during CE (see below).

The stellar models are computed at solar metallicity following \citealt{Asplund2009} for the solar abundance $Z_{\odot}=0.0142$ and their relative metal mass fractions.

Convection was modelled using the Ledoux criterion (\citealt{Ledoux1947}) within the standard mixing-length theory (\citealt{Bohm-Vitense1958}), with a mixing-length parameter $\alpha_{\mathrm{MLT}}=2$. Semiconvection was modelled according to \cite{Langer1983} with an efficiency parameter $\alpha_{\mathrm{sc}}=1$. For hydrogen-burning cores, we include step-overshooting extending the convective region by 0.10 pressure scale height at the convective boundary (but see Appendix \ref{sec:appVariations} where we discuss the impact of varying this number). For convective cores after the main-sequence, we include exponential overshooting (\citealt{Herwig2000}) with decay length of $f =$ 0.01. Thermohaline mixing is modeled as in \cite{Kippenhahn1980} with an efficiency parameter of $\alpha_{\mathrm{thermohaline}}=1$.

Nuclear reaction rates in \text{MESA} are from JINA REACLIB \citep{Cyburt2010}, NACRE \citep{Angulo1999} and additional tabulated weak reaction rates \citet{Fuller1985, Oda1994, Langanke2000}. We use the simple networks provided with \text{MESA} \texttt{basic.net} for H and He burning, which includes H$^1$, He$^3$, He$^4$, C$^{12}$, N$^{14}$, O$^{16}$, Ne$^{20}$ and Mg$^{24}$. As mentioned in the main text, our models run until core He-depletion, as failed-CE stripped stars produced from the merger of stars with inert carbon-oxygen cores are expected to be short-lived and therefore unlikely to contribute significantly to the overall population. 

Our models have no stellar winds. Firstly, for the considered mass range we expect radiation-driven winds to be weak on the Main Sequence (e.g. \citealt{Bjorklund2023}). Additionally, we adopt this approximation due to the low ($10^{-7}-10^{-9}\:M_{\odot}\:\mathrm{yr}^{-1}$) wind mass-loss rates measured for intermediate-mass stripped stars from the sample of \citealt{gotberg_stellar_2023}. We defer an exploration with a mass loss rate recipe for the different stripped products to future work. Additionally, we do not take into account stellar rotation or spin-orbit coupling.

\vspace{-0.35cm}
\section{CE formalism and merger products}\label{sec:appMethods}
\subsection{Onset of CE} 
We define the CE onset as the first step in the simulation in which either of these conditions occurs: 

\begin{itemize}
    \item[a)] L2-overflow: both stars overflow past the distance of their own outer Lagrangian point L2. We calculate the volume equivalent radii associated with the L2 point for the two components following Eq. (5.4) and Eq. (5.5) in \cite{PabloThesis2017}. We do not stop our MESA runs but rather let our simulations go further than this point. However, after the L2-overflow condition is hit, we do not implement any L2 overflow mass and angular momentum loss (see also below). In the vast majority of our systems, after this condition a) is hit, we find that condition b) (see below) is also hit shortly after. In our analysis, we simply flag the systems that would trigger the L2-overflow condition and assume that the merger happens shortly after that point.

    \item [b)] MT rate limit: the MT rate $\dot{M}$ becomes larger than 10 times the Kelvin-Helmholtz rate of the donor star, $\dot{M}_{\mathrm{KH,d}}$:
    \begin{equation}
    \dot{M}_{\mathrm{KH,d}} \equiv \frac{M}{t_{\mathrm{KH,d}}}
    \qquad
    \mathrm{with}\quad
    t_{\mathrm{KH,d}} \equiv \frac{G M^2}{R L}\:,
    \end{equation}
    where we calculate this quantity at each timestep, with $R$ being the radius, $L$ the luminosity, $M$ the mass of the star and $G=6.67 \times 10^{-8}\ \mathrm{cm^{3}\ g^{-1}\ s^{-2}}$.
    This is also the starting condition of our double core CE scheme (see below), regardless of whether or not condition a) is also holding.
    
\end{itemize} 
Based on the properties of the systems when one of these conditions is hit, we classify them as failed-CE stripped stars if they are additionally post-TAMS and there has been a previous stable MT episode; MS mergers otherwise (see the different types of MS mergers in Fig. \ref{fig:grid}). We further model the evolution of these binaries following the double core CE scheme described below. 


\vspace{-0.3cm}
\subsection{CE evolution} 
Once the MT rate limit (condition b) above) is hit, we start with our double core CE scheme. This is an extension of the method by \cite{Marchant2021} to the case of two stellar cores.
\paragraph{CE MT rate} We compute the MT rate during CE, $\dot{M}_{\mathrm{CE}}$, depending on the degree of RLOF of both components: at each timestep, we consider their overfilling factor, $\Delta_{j}$ for $j \in \{\mathrm{d,a}\}$, and the maximum value, $\Delta_\mathrm{max}$, as: 

\begin{equation}\label{eq:definition_MT_CE}
    \Delta_{j}\equiv \frac{R_j- R_{L,j}}{R_{L,j}}, \qquad
\Delta_{\max} = \max\!\left(\Delta_{\mathrm{a}}, \Delta_{\mathrm{d}}\right),
\end{equation}
where $R_{j}$ and $R_{L,j}$ are the radius and Roche-lobe radius of the accretor for $j=\mathrm{a}$ (donor for $j=\mathrm{d}$). If either star overfills its Roche lobe, the MT rate is set to $\dot{M}_{\mathrm{high}}=10\times\dot{M}_{\mathrm{KH,d}}$. Once both stars are detached, $\dot{M}_{\mathrm{CE}}$ is smoothly reduced to $\dot{M}_{\mathrm{low}}=10^{-8}\:\mathrm{M}_{\odot}\:\mathrm{yr}^{-1}$. Specifically: 

\begin{equation}\label{eq:Mdot_CE}
    \dot{M}_{\mathrm{CE}} =
\begin{cases}
\dot{M}_{\mathrm{high}}, & \Delta_{\max} > 0, \\[6pt]
\dot{M}_{\mathrm{low}},  & \Delta_{\max} < -0.02 \\[6pt]
\dot{M}_{\mathrm{high}}
\exp\!\left[
\displaystyle
\frac{\ln\!\left(\dot{M}_{\mathrm{high}}/\dot{M}_{\mathrm{low}}\right)}
{0.02}
\,\Delta_{\max}
\right], & -0.02 \le \Delta_{\max} \le 0 \, .
\end{cases}
\end{equation}
This prescription ensures a smooth transition between contact and detachment, and allows CE mass loss to persist until both stars are detached. Notice that the exponential form ensures that for $\Delta_{\mathrm{max}}\rightarrow0$ (i.e., when at least one star is about to fill its own Roche lobe), the CE mass loss rate $\dot{M}_{\mathrm{CE}}\rightarrow\dot{M}_{\mathrm{high}}$; vice versa, for $\Delta_{\mathrm{max}}\rightarrow -0.02$ (i.e., both stars are detached), then $\dot{M}_{\mathrm{CE}}\rightarrow\dot{M}_{\mathrm{low}}$.

Additionally, we allow both binary components to lose mass. The total CE mass-loss rate $\dot{M}_{\rm CE}$ is distributed between the donor and accretor according to their degree of Roche-lobe overfilling. For each component $j \in \{\mathrm{d,a}\}$, we define 
\begin{equation}
    \delta_j = \frac{R_j - R_{L,j}}{\min(R_j, R_{L,j})}\:,
\end{equation}
and assign a weighting factor $f_j = \exp(\delta_j)$. The resulting mass-loss rates of the donor and accretor are then given by
\begin{equation}
    \dot{M}_\mathrm{d} = \dot{M}_{\rm CE}\,\frac{f_\mathrm{d}}{f_\mathrm{d} + f_\mathrm{a}},
\qquad
\dot{M}_\mathrm{a} = \dot{M}_{\rm CE}\,\frac{f_\mathrm{a}}{f_\mathrm{d} + f_\mathrm{a}},
\end{equation}
such that the component that more strongly overfills its Roche lobe contributes a larger fraction of the envelope mass loss. This prescription allows both stars to participate in the envelope ejection during CE evolution.


\vspace{-0.3cm}
\paragraph{Energy formalism for the double-core CE} Our double-core CE scheme includes both the donor and accretor envelopes. At the onset of CE, we compute and store the mass-coordinate-dependent internal and gravitational energies of both stars, $U(m)$ and $\Omega(m)$, corrected for recombination energies. These are later used to compute the amount of envelope binding energy $E_{\mathrm{bind}}$ that has been removed at each timestep $t$ as the stars are stripped to a mass coordinate $m_t$. \\
During CE evolution, the binary orbital separation is updated using the energy formalism: for each star stripped to $m_t$, the envelope binding energy is given by
\begin{equation}
E_{\mathrm{bind}} = \alpha_{\mathrm{th}}\,U(m_t) + \Omega(m_t),
\end{equation}
where $\alpha_{\mathrm{th}}$ is the CE thermal efficiency factor, for which we assume $\alpha_{\mathrm{th}}=1$, $U(m_t)$ and $\Omega(m_t)$ are the internal and gravitational energies, respectively, interpolated from the stored profiles $U(m)$ and $\Omega(m)$ at CE onset. 
The orbital separation at each timestep, $a_t$, is updated by solving the energy equation including both stellar envelopes,
\begin{equation}
\alpha_{\mathrm{CE}} \left( E_{\mathrm{orb,}t} - E_{\mathrm{orb,i}} \right) = - \left( E_{\mathrm{bind,a}} + E_{\mathrm{bind,d}} \right),
\end{equation}
where $\alpha_{\mathrm{CE}}=1$ is the CE efficiency factor, $E_{\mathrm{orb,i}} = - G M_{\mathrm{d,i}} M_{\mathrm{a,i}} / (2 a_{\mathrm{i}})$ is the initial orbital energy, and $E_{\mathrm{bind,a}}$ ($E_{\mathrm{bind,d}}$) is the binding energy of the accretor (donor). 

\vspace{-0.3cm}
\subsection{End of CE and the merger product}
The system survives the CE as a binary if both the donor and the accretor a) detach from their Roche lobes, i.e. $\Delta_\mathrm{a},\:\Delta_{\mathrm{d}}<-0.02$; b) maintain their radius such that $R<R_L$ for longer than 1\% of the Kelvinh-Helmholtz timescale of the donor, $t_{\mathrm{KH,d}}$. If either condition is met, the CE phase is terminated and binary evolution resumes.\\
Otherwise, the CE is terminated as a merger if one of the following conditions is met:
\begin{enumerate}
    \item[a)] The donor is stripped to a chemically homogeneous core, defined by
\begin{equation}
\left| X_{\mathrm{H,center}} - X_{\mathrm{H,surface}} \right| < 0.01 \hspace{0.25cm}\mathrm{and}
\end{equation}
\[\left| X_{\mathrm{He,center}} - X_{\mathrm{He,surface}} \right| < 0.01\:,\]
where $X$ represents mass fractions of the considered element.

\item[b)] After CE stripping, the core, at mass coordinate $m_{\mathrm{core}}$, of either the donor or the accretor, would have a radius $R_{\mathrm{core}}$ such that it would overflow the Roche lobe:
\begin{equation}
R_{\mathrm{core}} > R_L(m_{\mathrm{core}})\:.
\end{equation}
\item[c)] The final orbital period, after CE, falls below 5 hours.
\end{enumerate}
We build the merger product by combining the mass shells of both stars into a total mass $m_{\mathrm{merger}} = m_1 + m_2$, with more evolved (hydrogen-poor) material placed at smaller mass coordinates. The merged star is then relaxed in composition before further evolution till He depletion.

\vspace{-0.3cm}
\section{A selection of our grids}\label{sec:appGrids}
As discussed in the main text, after $M_{1,\mathrm{i}}\geq10\:M_{\odot}$ we stop resolving the failed-CE stripped stars produced by a first stage of case B MT, due to the region becoming closer and closer to $q_{\mathrm{i}}\simeq 1$. On the other hand, case AB MT systems are still present at our resolution for $M_{1,\mathrm{i}}=10\:M_{\odot}$, but we stop resolving them at $M_{1,\mathrm{i}}>18\:M_{\odot}$, as shown below in Fig. \ref{fig:grid_18Msun}. We note that, in this case, the region below TAMS becomes instead populated by RLOF stripped stars.

Additionally, to support Fig. \ref{fig:shapes}, we show in Fig. \ref{fig:grids_figure_3} the full model grids of primary masses $M_{1,\mathrm{i}}=4,\:8,\:10\:M_{\odot}$. This is only a selection of the full set of computed grids; the remaining figures will be available at the time of publication on the dedicated \href{https://doi.org/10.5281/zenodo.17424084}{Zenodo} entry for this work.

Interestingly, we note that we predict only a small subset of our models to survive the CE episode as a binary, at least until core-He depletion (at which we terminate our simulations). We do not study these further in this work. Notably, none of the systems that initiate a post-TAMS inverse MT survives the CE episode; they all evolve into failed-CE stripped stars.

\vspace{1cm}
\begin{figure}[h!]
   \centering
   \vspace{-0.4cm}
   \includegraphics[width=\hsize]{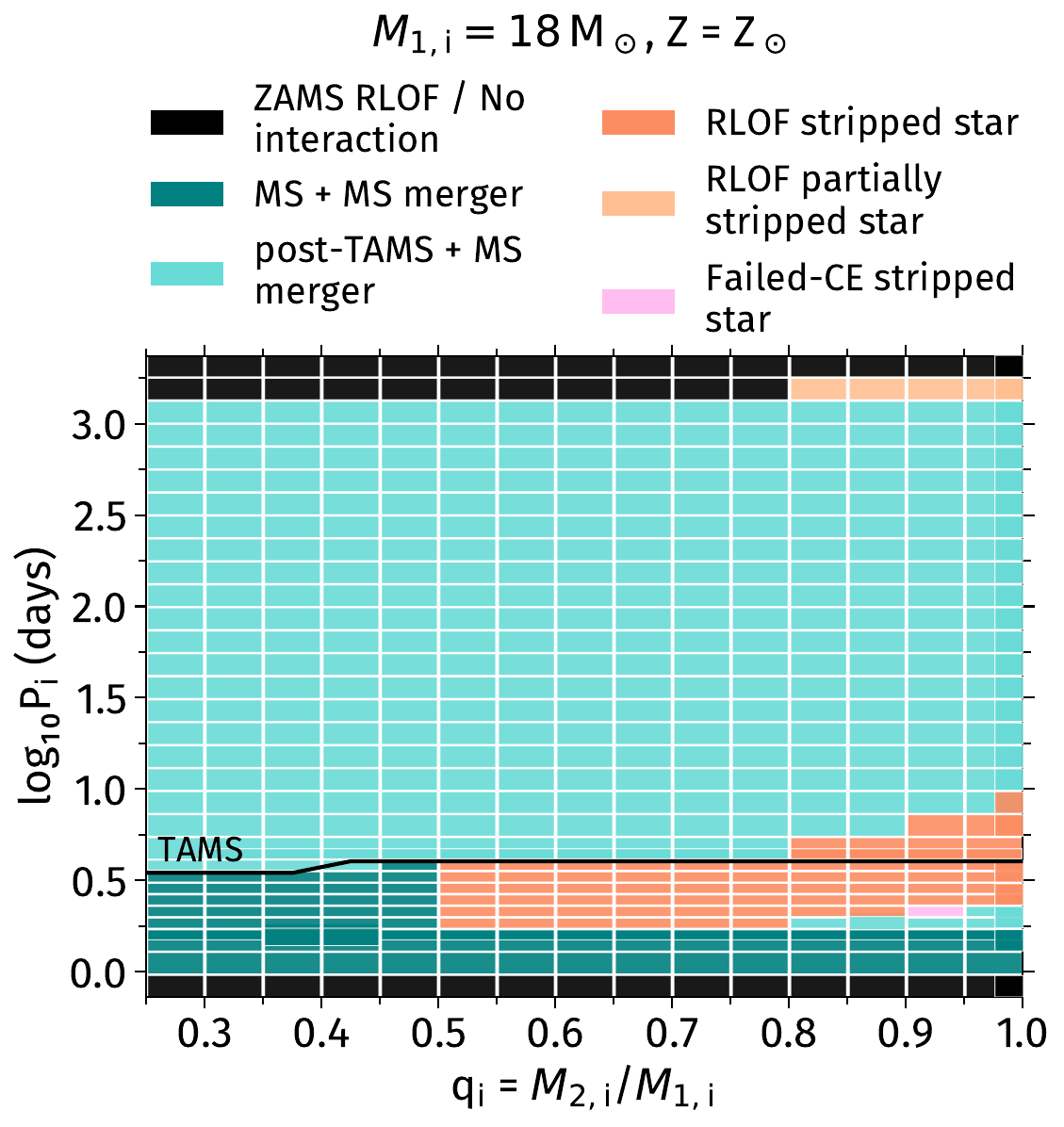}
      \caption{Grid of the primary mass $M_{1,\mathrm{i}}=18\:M_{\odot}$, with the same color coding and conventions of Fig. \ref{fig:grid}. At our resolution, we stop resolving the failed-CE parameter space for $M_{1,\mathrm{i}}>18\:M_{\odot}$.}
         \label{fig:grid_18Msun} 
   \end{figure}

 \begin{figure}[h!]
   \centering
   \includegraphics[width=\hsize]{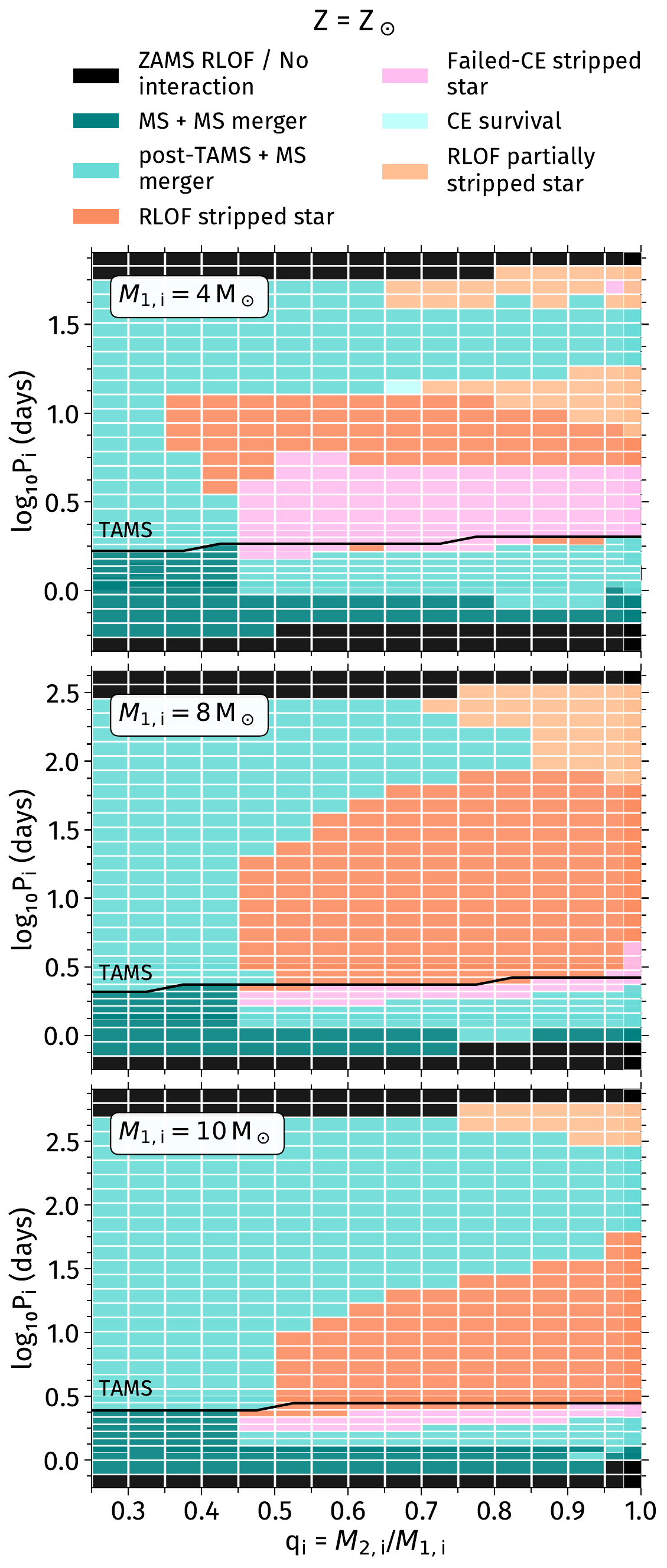}
      \caption{Grids for primary masses $M_{1,\mathrm{i}}=4,\:8,\:10\:M_{\odot}$ considered in Fig. \ref{fig:shapes}, with the same conventions of Fig. \ref{fig:grid}. With respect to Fig. \ref{fig:grid}, one additional possible outcome is shown here: CE survival, i.e. systems surviving as binaries after CE. This happens for a small subset of our models and we do not study it further.}
         \label{fig:grids_figure_3} 
   \end{figure}
\FloatBarrier

\vspace{-0.3cm}
\section{Who is the fastest? The shape of the failed-CE parameter space}\label{sec:appContours}
As discussed in the main text, the failed-CE stripped star outcome from an initial case B MT episode is limited from above at initial orbital periods that depend on the initial mass ratio. This gives rise to the "triangular" shape above TAMS that we see in Fig. \ref{fig:grid}, and shapes the parameter space of failed-CE mergers at different masses as shown in Fig. \ref{fig:shapes}. 

\begin{figure}[h!]
   \centering
   \includegraphics[width=\linewidth]{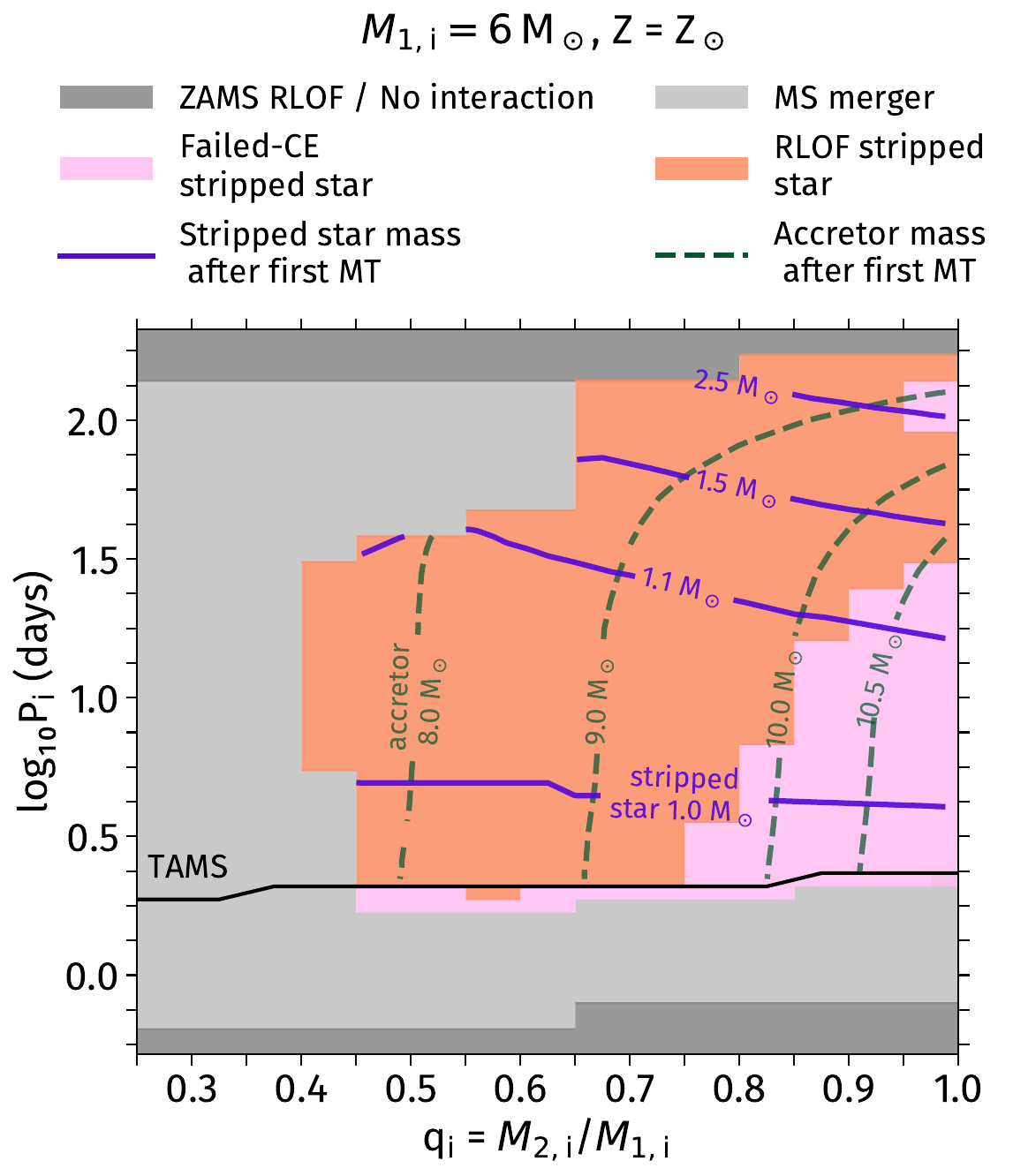}
      \caption{Same model grid as Fig. \ref{fig:grid}, but with a different color coding to better highlight the RLOF and Failed-CE stripped stars outcomes. We refer to any merger happening with at least one MS star as MS merger (collectively describing both the post-TAMS + MS and MS + MS merger outcomes). Iso-contours of fixed accretor (stripped star) mass after the first stable MT episode are shown in dashed green (solid violet) lines. }
         \label{fig:ML_relationship} 
   \end{figure}

In Fig. \ref{fig:ML_relationship}, we illustrate such a behavior in terms of the masses of the binary components after the first stable MT episode. At a fixed initial mass ratio, longer initial periods result in partial envelope stripping of the primary after the first MT episode, thereby producing more massive RLOF stripped stars that will evolve on a faster timescale. This effect is acting against the failed-CE stripped star outcome, as the secondary has less time to initiate the CE before the stripped primary depletes He in its core. However, for initial mass ratios sufficiently close to unity, the secondary star at detachment can be heavy enough to \textit{be the fastest}, but this will depend on how much mass is actually transferred from the primary (we refer to this as effect 1) in the main text), and on its mass-luminosity relationship, which dictates how much faster a star effectively becomes after accreting mass (effect 2) in the main text).

\vspace{-0.35cm}
\section{Full evolutionary tracks of Figure 4}\label{sec:appHR}
\begin{figure}[h]
\centering
\includegraphics[width=0.925\hsize]{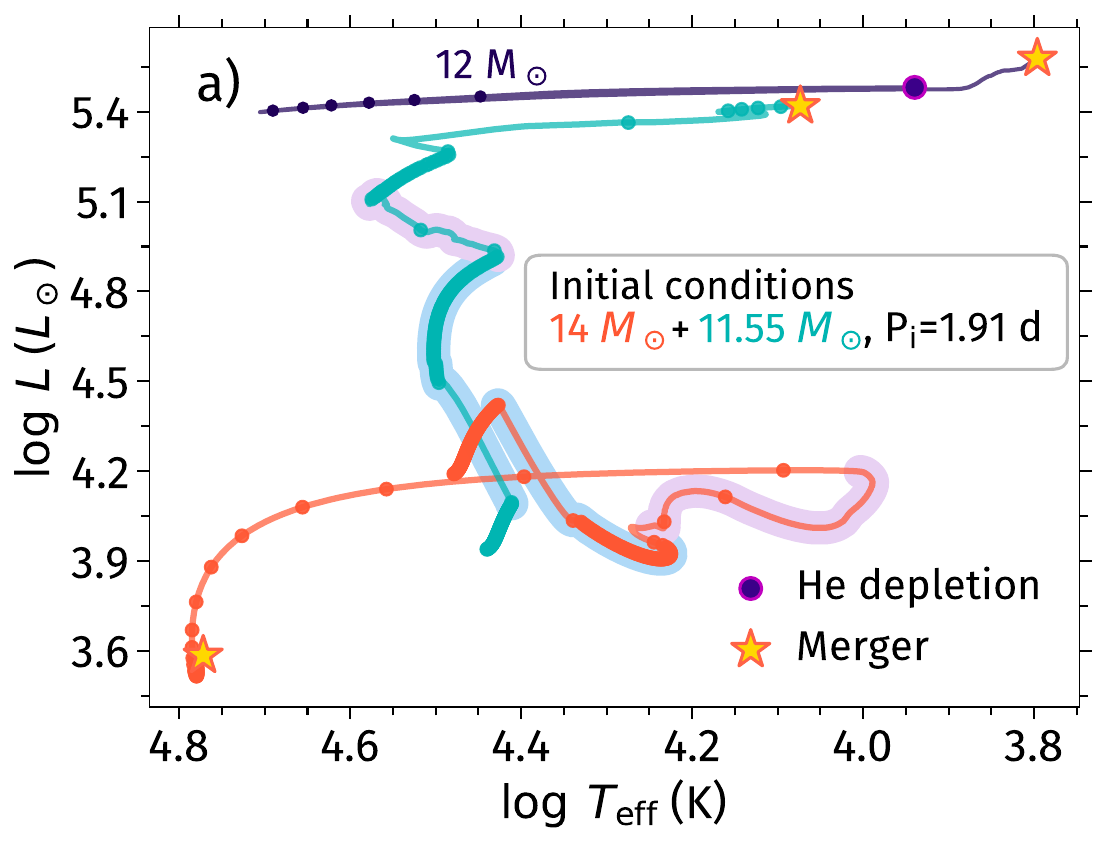}
\includegraphics[width=0.925\hsize]{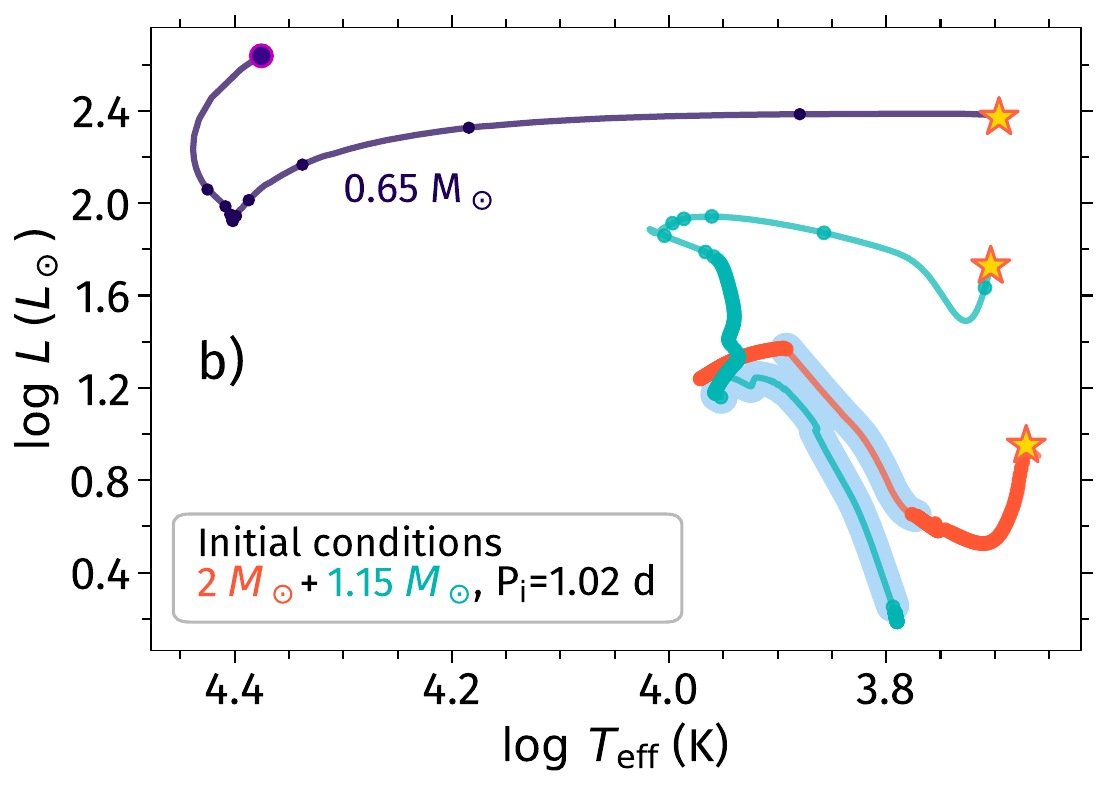}
\includegraphics[width=0.925\hsize]{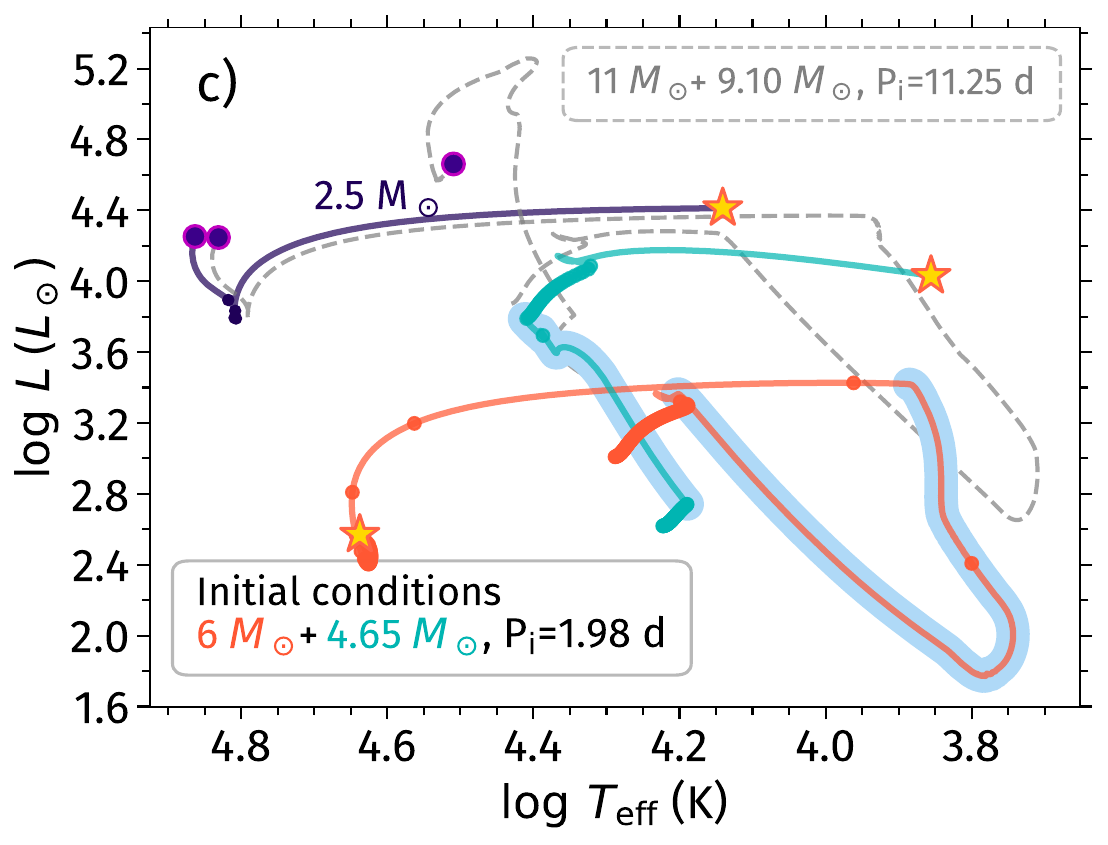}
\caption{a) HR diagram of a $12\:M_{\odot}$ failed-CE stripped star model (purple line), up to He depletion (purple marker). This is the product of a binary with 14 $M_{\odot}$ and 11.55 $M_{\odot}$ and initial orbital period $P_{\mathrm{i}}=1.91\:\mathrm{days}$. We show the evolution of the progenitor primary (secondary) in orange (cyan), up to the failed-CE merger onset (star marker). Scatter points along the tracks are spaced by 0.05 Myrs, and MT episodes (case A and AB) are highlighted as colored areas; b) Failed-CE stripped star of mass $0.65\:M_{\odot}$, produced by a binary of 2 $M_{\odot}$ and 1.15 $M_{\odot}$ and initial orbital period $P_{\mathrm{i}}=1.02\:\mathrm{days}$. Scatter points along the tracks are every 5 Myrs; c) Failed-CE stripped star of mass $2.5\:M_{\odot}$, formed by a binary of 6 $M_{\odot}$ and 4.65 $M_{\odot}$ and initial orbital period $P_{\mathrm{i}}=1.98\:\mathrm{days}$. Scatter points along the tracks are every 0.5 Myrs. We add the evolution (dashed gray) of a binary with an 11 $M_{\odot}$ primary that evolves into a $2.5\:M_{\odot}$ RLOF stripped star, when in orbit with a 9.10 $M_{\odot}$ secondary at $P_{\mathrm{i}}=11.25$ days.
}
\label{fig:HR_app}
\end{figure}

To support Fig. \ref{fig:HR}, we show in Fig. \ref{fig:HR_app} the evolutionary tracks of the failed-CE stripped stars with their progenitor binaries.

\vspace{-0.38cm}
\section{Parameter variations}\label{sec:appVariations}
\begin{figure}[t]
   \centering
   \includegraphics[width=\hsize]{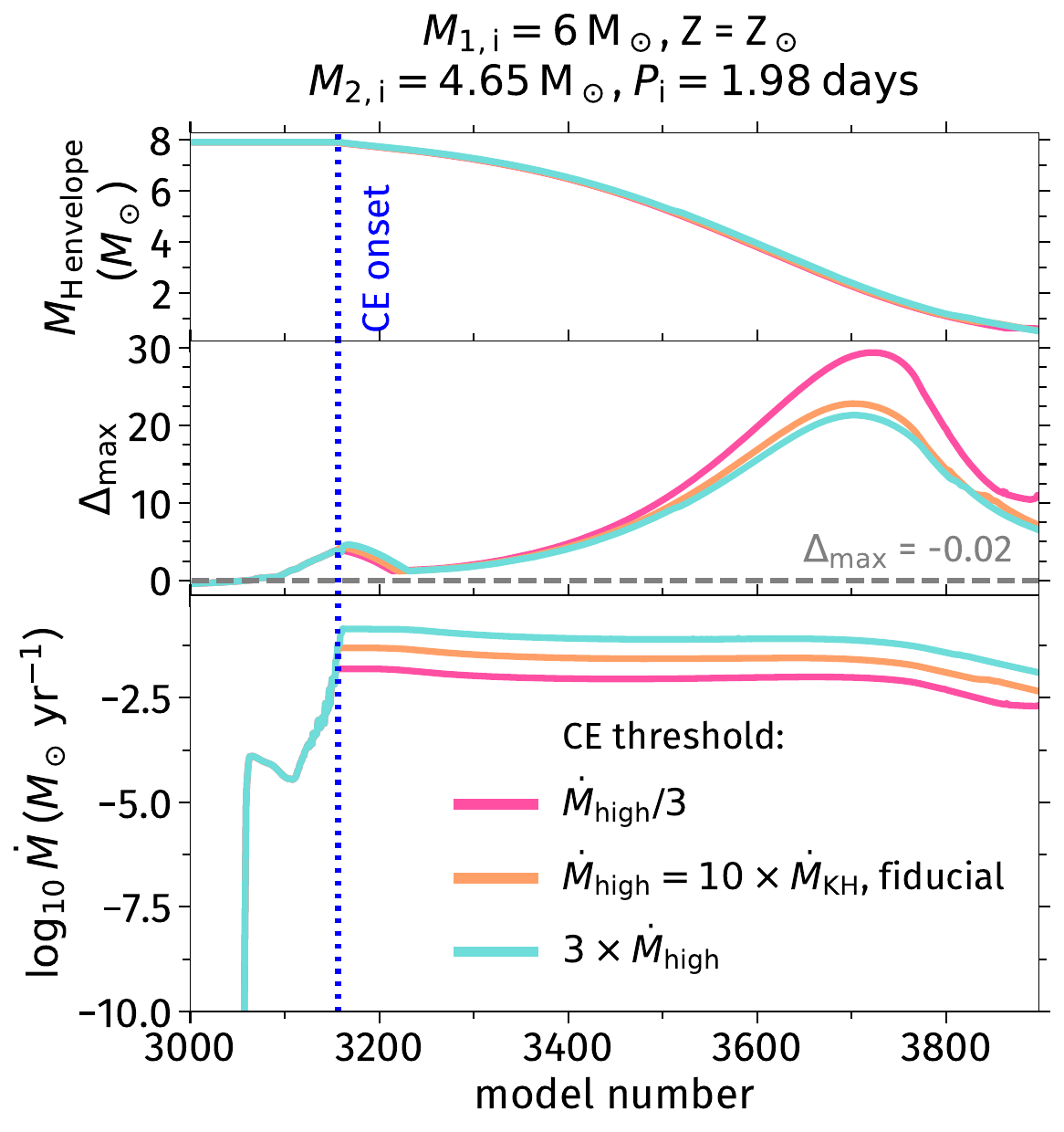}
      \caption{Example of MT history during the failed CE episode for our system with initial masses $6\:M_{\odot}$ and $4.65\:M_{\odot}$ at initial orbital period $P_{\mathrm{i}}=1.98$ days (the same represented in panel c) of Fig. \ref{fig:HR_app}, and in Fig. \ref{fig:HR} as progenitor of the $2.5\:M_{\odot}$ merger). As described in Appendix \ref{sec:appMethods}, we switch on our double-core CE scheme when different CE thresholds are reached for the first time; after this, we adopt the recipe of Eq. \ref{eq:Mdot_CE} based on $\Delta_{\mathrm{max}}$.}
         \label{fig:MKH_threshold}
   \end{figure}

Our results assume a fixed fiducial threshold for CE at $\dot{M}_{\mathrm{high}}=10\times\dot{M}_{\mathrm{KH,d}}$, and then adapts the MT rate based on the recipe described in Eq. \ref{eq:Mdot_CE}. We performed a test (see Fig. \ref{fig:MKH_threshold}), relaxing the CE threshold within one order of magnitude and found that the remaining hydrogen envelope may vary by 10\%, which is not negligible. While this does not affect our conclusions, we note that it will pose a significant effect in population predictions.

\begin{figure}[t]
   \centering
   \includegraphics[width=\hsize]{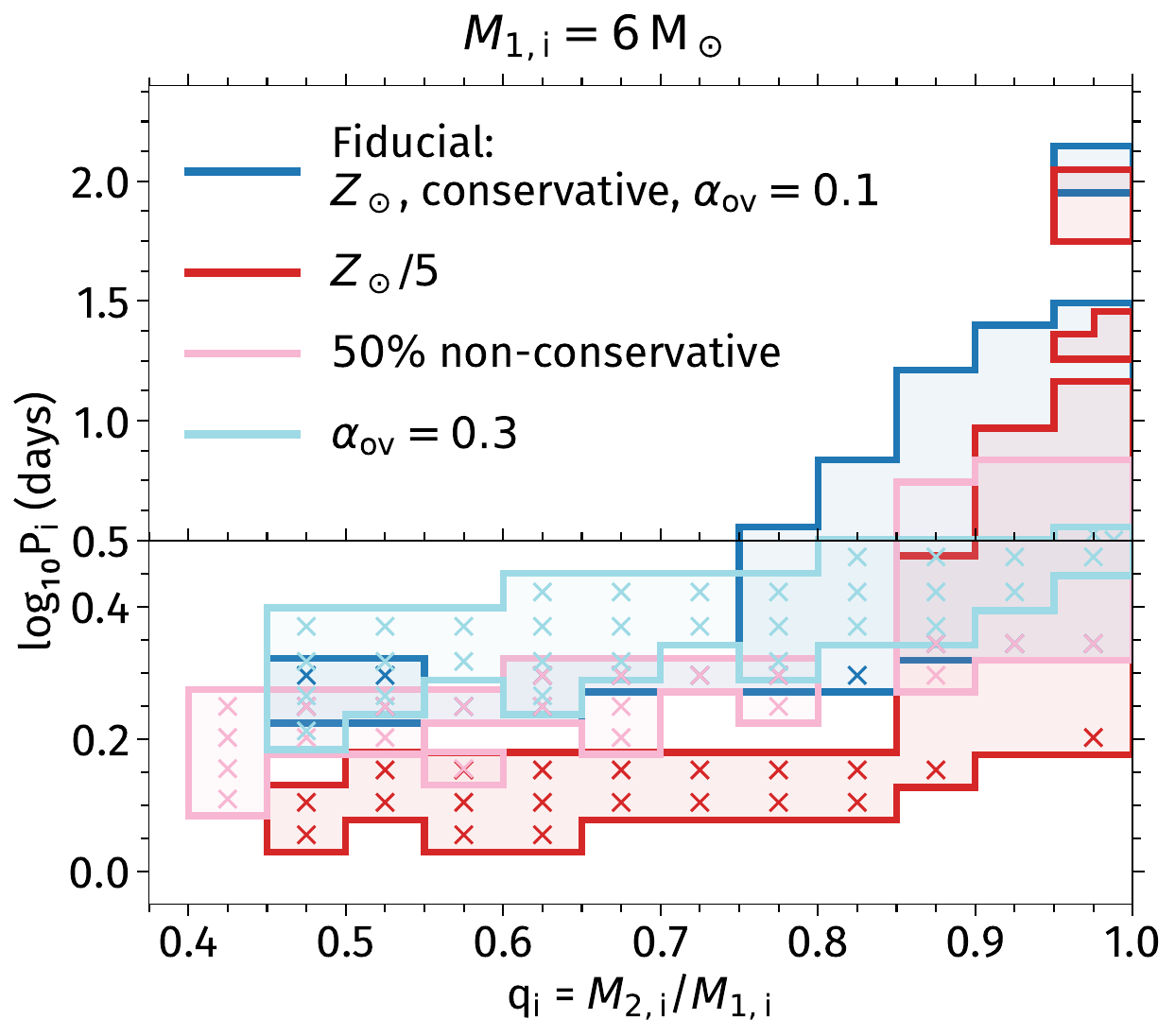}
      \caption{Parameter space in initial period $P_\mathrm{i}$ and mass ratio $q_\mathrm{i}$ for failed-CE stripped stars for an initial primary mass $M_{1,\:\mathrm{i}}=6\:M_{\odot}$. Different colors show the variations with respect to our fiducial model with solar metallicity, fully conservative MT, and convective overshooting parameter $\alpha_{\mathrm{ov}}=0.1$. As in Fig. \ref{fig:shapes}, systems undergoing case A MT in the first interaction are indicated with x scatter points.}
         \label{fig:variation}
   \end{figure}

\begin{figure}[t]
   \centering
   \includegraphics[width=\hsize]{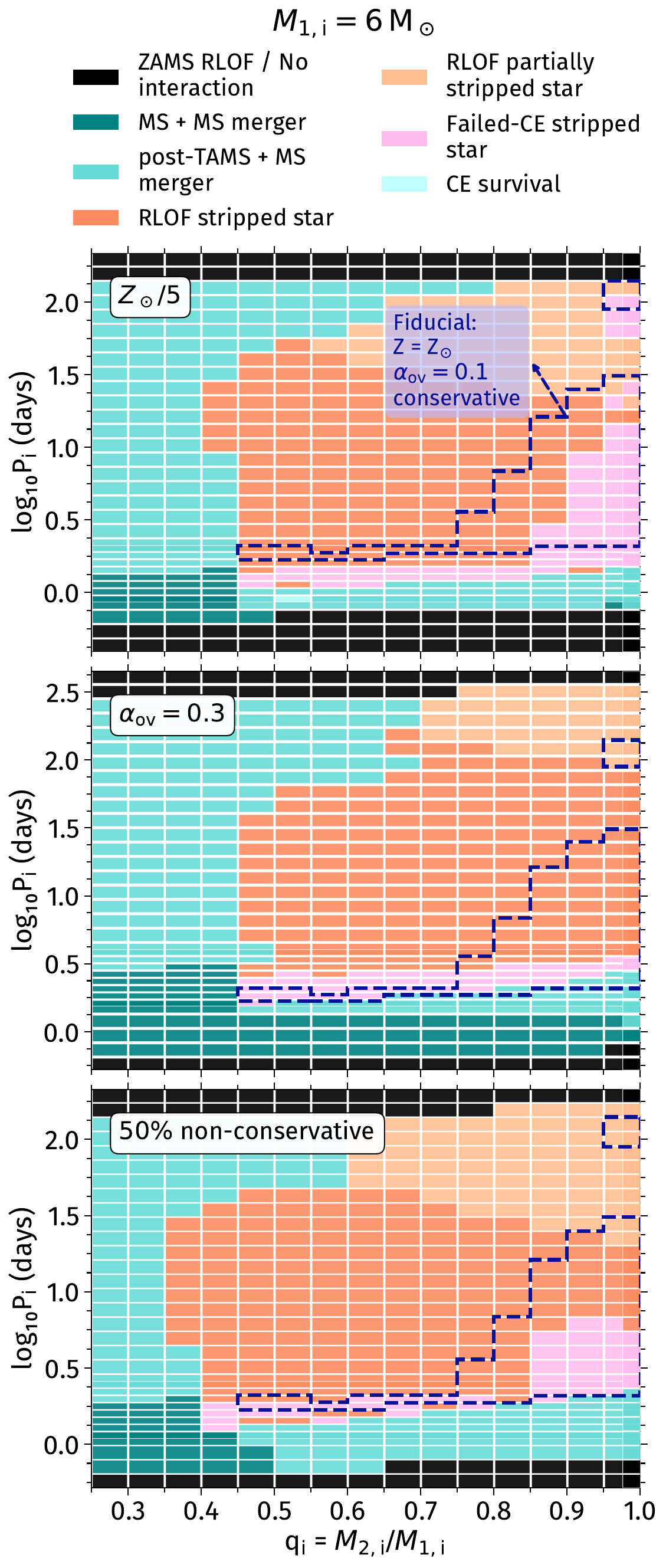}
      \caption{Model grids of initial primary mass $M_{1,\mathrm{i}}=6\:M_{\odot}$ considered in Fig. \ref{fig:variation}, with the same color coding and conventions of Fig. \ref{fig:grid} and Fig. \ref{fig:grids_figure_3}. The three panels show the three parameter variations we have explored, and we overplot the failed-CE stripped star parameter space of the fiducial model grid (that of Fig. \ref{fig:grid}) as an indigo-dashed region.}
         \label{fig:grids_parameter} 
   \end{figure}

Our results are computed at solar metallicity. However, the observed sample of \citet{gotberg_stellar_2023} is drawn from the Magellanic Clouds, and metallicity is expected to affect both the predicted fraction of stripped stars \citep{HovisAfflerbach2025} and the parameter space for (un)stable reverse MT \citep[e.g.][]{Pauli2024,Briel2025}. Lower metallicity modifies the expansion history of the binary components, shifting the failed-CE stripped star parameter space toward shorter initial periods. We verify this expectation using one grid at $Z=Z_{\odot}/5$, which posed 24\% reduction of the failed-CE mergers parameter space with respect to $Z=Z_{\odot}$. The results of this experiment are presented in Fig. \ref{fig:variation}, where the failed-CE parameter space at $Z_{\odot}/5$ is compared to the fiducial case at galactic metallicity; for completeness, we also show in Fig. \ref{fig:grids_parameter} (top panel) the full models grid. We defer a systematic exploration of metallicity to future work. 

Convective overshooting is also expected to affect stellar expansion. In Fig. \ref{fig:variation} and Fig. \ref{fig:grids_parameter} (middle panel) we show the results from a test with enhanced overshooting, extending the $\alpha_{\mathrm{ov}}$ parameter to 0.3 pressure scale heights compared to the fiducial value 0.1. We find that stronger overshooting accelerates the primary evolution and suppresses the failed-CE stripped star parameter space by 64\%, especially at longer periods. Although we performed only one test, we can see that a higher overshooting parameter acts to produce a parameter space that mimicks that of a higher initial primary mass, in which the preference for $q_{\mathrm{i}}\simeq 1$ becomes stronger due to the effects described in the main text.

Finally, the efficiency of MT will have an impact on our results, as lower accretion efficiency would limit the rejuvenation of the secondary and the growth of its Roche lobe. Our fiducial models assume fully conservative MT, motivated by recent evidence that conservative MT is the most likely formation channel for sdO + Be binaries \citep{Lechien2025}. We have carried on an experiment with an example grid computed with 50\% MT efficiency, which shows that longer-period systems are suppressed due to less mass being accreted from the secondary, with an overall suppression of the failed-CE stripped star parameter space of 50\%. These results are showcased in Fig. \ref{fig:variation} and Fig. \ref{fig:grids_parameter} (bottom panel). Additionally, our models do not include rotation, which may further affect MT efficiency and the stability of inverse MT onto the stripped primary \citep{Briel2025}.

%

%

%

\end{appendix}
\end{document}